\documentclass[12pt,letterpaper]{article}

\usepackage{epsfig}
\usepackage{amsmath}
\usepackage{amssymb}
\usepackage{amsthm}
\usepackage{indentfirst}
\usepackage{xspace}
\usepackage{multirow}
\usepackage{hyperref}
\usepackage{xcolor}
\definecolor{darkred}{rgb}{0.5,0.15,0.15}
\hypersetup{colorlinks=true,urlcolor=darkred,linkcolor=darkred,citecolor=darkred}
\usepackage{verbatim}
\usepackage[letterpaper,margin=1in,headheight=15pt]{geometry}
\usepackage{mathpazo}
\usepackage{tikz-cd}
\usepackage{booktabs}
\usepackage{framed}
\definecolor{shadecolor}{rgb}{0.85,0.85,0.85}

\usepackage[bibstyle=authoryear-comp,labelyear=false,defernumbers=true,maxnames=20,firstinits=true, uniquename=init,dashed=false,backend=biber]{biblatex}

\DeclareNameAlias{sortname}{first-last}

\DeclareFieldFormat{url}{\url{#1}}
\DeclareFieldFormat[article]{pages}{#1}
\DeclareFieldFormat[inproceedings]{pages}{\lowercase{pp.}#1}
\DeclareFieldFormat[incollection]{pages}{\lowercase{pp.}#1}
\DeclareFieldFormat[article]{volume}{\textbf{#1}}
\DeclareFieldFormat[article]{number}{(#1)}
\DeclareFieldFormat[article]{title}{\MakeCapital{#1}}
\DeclareFieldFormat[inproceedings]{title}{#1}
\DeclareFieldFormat{shorthandwidth}{#1}

\DeclareBibliographyDriver{article}{%
  \usebibmacro{bibindex}%
  \usebibmacro{begentry}%
  \usebibmacro{author/editor}%
  \setunit{\labelnamepunct}\newblock
  \MakeSentenceCase{\usebibmacro{title}}%
  \newunit
  \printlist{language}%
  \newunit\newblock
  \usebibmacro{byauthor}%
  \newunit\newblock
  \usebibmacro{byeditor+others}%
  \newunit\newblock
  \printfield{version}%
  \newunit\newblock
  \usebibmacro{journal+issuetitle}%
  \newunit\newblock
  \printfield{note}%
  \setunit{\bibpagespunct}%
  \printfield{pages}
  \newunit\newblock
  \usebibmacro{eprint}
  \newunit\newblock
  \printfield{addendum}%
  \newunit\newblock
  \usebibmacro{pageref}%
  \usebibmacro{finentry}}

\renewbibmacro*{journal+issuetitle}{%
  \usebibmacro{journal}%
  \setunit*{\addspace}%
  \iffieldundef{series}
    {}
    {\newunit
     \printfield{series}%
     \setunit{\addspace}}%
  \printfield{volume}%
  \printfield{number}%
  \setunit{\addcomma\space}%
  \printfield{eid}%
  \setunit{\addspace}%
  \usebibmacro{issue+date}%
  \newunit\newblock
  \usebibmacro{issue}%
  \newunit}

\def\makebibcategory#1#2{\DeclareBibliographyCategory{#1}\defbibheading{#1}{\section*{#2}}}
\makebibcategory{books}{Books}
\makebibcategory{papers}{Refereed research papers}
\makebibcategory{chapters}{Book chapters}
\makebibcategory{conferences}{Papers in conference proceedings}
\makebibcategory{techreports}{Unpublished working papers}
\makebibcategory{bookreviews}{Book reviews}
\makebibcategory{editorials}{Editorials}
\makebibcategory{phd}{PhD thesis}
\makebibcategory{subpapers}{Submitted papers}
\makebibcategory{curpapers}{Current projects}

\renewcommand*{\bibitem}{\addtocounter{papers}{1}\item \mbox{}\hskip-0.85cm\hbox to 0.85cm{\hfill\arabic{papers}.~~}}
\defbibenvironment{bibliography}
{\list{}
  {\setlength{\leftmargin}{\bibhang}%
   \setlength{\itemsep}{\bibitemsep}%
   \setlength{\parsep}{\bibparsep}}}
{\endlist}
{\bibitem}

\newcounter{papers}
\newcounter{sumpapers}

\numberwithin{equation}{section}

\newcommand{\fg}{{\mathfrak g}}

\newcommand{\cC}{\ensuremath{\mathcal C}}
\newcommand{\cG}{\ensuremath{\mathcal G}}
\newcommand{\cB}{\ensuremath{\mathcal B}}
\newcommand{\cL}{\ensuremath{\mathcal L}}

\newcommand{\cK}{\ensuremath{\mathcal K}}

\newcommand{\cM}{\ensuremath{\mathcal M}}
\newcommand{\cO}{\ensuremath{\mathcal O}}

\newcommand{\cA}{\ensuremath{\mathcal A}}
\newcommand{\cI}{\ensuremath{\mathcal I}}

\newcommand{\R}{\ensuremath{\mathbb R}}
\newcommand{\C}{\ensuremath{\mathbb C}}
\newcommand{\PP}{\ensuremath{\mathbb P}}
\newcommand{\Z}{\ensuremath{\mathbb Z}}

\newcommand{\bS}{\ensuremath{\mathbb S}}

\newcommand{\N}{{\mathcal N}}

\newcommand{\hk}{hyperk\"ahler\xspace}

\newcommand{\breg}{{\cB_{\mathrm{reg}}}}
\newcommand{\bsing}{{\cB_{\mathrm{sing}}}}

\newcommand{\ab}{\mathrm{ab}}

\newcommand{\IP}[1]{\langle#1\rangle}

\newcommand{\ti}[1]{\textit{#1}}

\newcommand{\fro}{\overline{\underline{\Omega}}}

\newcommand{\hb}{harmonic bundle}

\DeclareMathOperator{\Tr}{Tr}
\DeclareMathOperator{\End}{End}
\DeclareMathOperator{\Hom}{Hom}
\DeclareMathOperator{\Aut}{Aut}

\bibliography{hitchin-systems}

\begin{document}

\title{Hitchin systems in $\N=2$ field theory}
\author{Andrew Neitzke}

\maketitle

\setcounter{page}{1}

\section{Introduction}

This note is a short review of the way Hitchin systems appear in four-dimensional 
$\N=2$ supersymmetric field theory.

The literature on the Hitchin system and its role in quantum field theory is a vast one.
Restricting attention just to the role of Hitchin systems 
in $\N=2$ supersymmetric field theory (thus neglecting such fascinating topics as $T$-duality 
on the Hitchin fibration and its relation to the geometric
Langlands program \cite{mlh,Harvey:1995tg,geom-lang-n4,Gaiotto:2009hg}, 
the use of Hitchin systems in $\N=4$ super Yang-Mills \cite{Alday:2009dv,Alday:2009yn,Alday:2010vh}, 
the role of Higgs bundles in F-theory \cite{Donagi2009}, \dots) cuts things down somewhat
but still leaves an enormous pool of papers and topics from which to choose. 
In this article I focus on the points with which I am most personally familiar.
In particular, although this review is meant for a special volume devoted to the AGT correspondence,
I will have very little to say about that.  This is not because I think there is nothing to say ---
on the contrary, works such as \cite{Nekrasov:2010ka,Teschner2011} have demonstrated that there
clearly is --- but because I do not know precisely what to say.

In one sentence, the relation between $\N=2$ theories and Hitchin systems is that
the Hitchin system arises as the \ti{moduli space} of the $\N=2$ theory compactified on
a circle.
My aim in this note is to explain a dictionary between various aspects of the field theory 
(its Coulomb branch, its
line defects, its surface defects, \dots) and their manifestations in the Hitchin system
(the Hitchin base, some distinguished holomorphic functions, some distinguished hyperholomorphic
bundles, \dots), along with a few ways in which this dictionary gives insight into aspects
of the Hitchin system.

My perspective on this subject has been heavily influenced by a long and enjoyable collaboration
with Davide Gaiotto and Greg Moore.  It is a pleasure to thank them for this collaboration
and for the many things that they have taught me.  This work is supported by NSF grant 1151693.

\medskip

In \S\ref{sec:n=2-generalities} we review general facts about $\N=2$ theories, their
relation to integrable systems and \hk geometry, and line and surface defects therein.  
The Hitchin system does not appear explicitly in this section.
In \S\ref{sec:class-s} we specialize to the 
case of theories of class $S$; this is the class of $\N=2$ 
theories most directly related to Hitchin systems.
Finally, in \S\ref{sec:hitchin-systems} we give some general 
background on the Hitchin system, divorced
from its role in physics; this section could in principle be read on its own,
but is mainly intended as a reference for selected facts which we will need in the other sections.

Each subsection of \S\ref{sec:n=2-generalities} and \S\ref{sec:class-s} is preceded
by a brief slogan.  It may be worth reading all the slogans first, to get an idea of
what is going on here.

\section{$\N=2$ theories and their circle compactification} \label{sec:n=2-generalities}

In this section we briefly review some facts about $\N=2$ theories $T$ in $4$ dimensions, and their 
compactification on a circle $R$ to give theories $T[R]$ in $3$ dimensions.

We will describe only general 
features here, without specializing to any particular theory $T$; in the next section we will
explain how all of these general phenomena are realized in the special case of theories of class $S$.

\subsection{$\N=2$ theories in the IR and integrable systems} \label{sec:integrable-system}

\begin{shaded}
Any $\N=2$ theory gives rise to a complex integrable system.
\end{shaded}

Consider an $\N=2$ supersymmetric theory $T$ in $4$ dimensions.
Let $\cB$ denote the Coulomb branch.
$\cB$ consists of an open ``regular locus'' $\breg$
plus a ``discriminant locus'' $\bsing$.

The IR physics in vacua labeled by points $u \in \breg$
is governed by pure abelian $\N=2$ gauge theory, with gauge group $U(1)^r$, where $r = \dim_\C \cB$.
Locally around any point $u \in \breg$, this IR theory can be described in terms of classical
fields, namely $r$ $\N=2$ vector multiplets.
The bosonic field content is thus $r$ complex scalars and $r$ abelian gauge fields.
However, there is generally no single Lagrangian that describes the IR theory globally
on $\breg$:  rather, we must use different Lagrangians in different patches of $\breg$,
related to one another by electric/magnetic duality transformations.
This story was first worked out in \cite{Seiberg:1994aj,Seiberg:1994rs}.
Although we cannot write a single Lagrangian that describes the theory globally, there is a
single geometric object from which all the local Lagrangians can be derived
\cite{Donagi:1997sr,Donagi:1996cf,Martinec:1995by}.
This object is a \ti{complex integrable system}:  a holomorphic symplectic manifold $\cI'$,
with a projection $\pi: \cI' \to \breg$, such that the fibers $\cI'_u = \pi^{-1}(u)$ are compact complex Lagrangian tori, of complex
dimension $r$.
One has the following dictionary:
\begin{center}
\begin{tabular}{cc} \toprule
Fiber of $\cI$ over $u \in \breg$ & IR physics at $u \in \breg$ \\ \midrule
 $H_1(\cI'_u, \Z)$  &  EM charge lattice $\Gamma_u$ \\
 polarization of $\cI'_u$ & DSZ pairing on $\Gamma_u$ \\
 symplectic basis of $H_1(\cI'_u, \Z)$ & electric-magnetic splitting \\
 automorphisms of $\cI'_u$  & EM duality group ($\simeq Sp(2r,\Z)$)\\
 period matrix of $\cI'_u$  & matrix of EM gauge couplings \\
 point of $\cI'_u$ & EM holonomies around surface defect \\
 \bottomrule
\end{tabular}
\end{center}


So far we have been discussing the IR physics at points $u$ in the regular locus $\breg$.
At $u \in \bsing$ the simple description of the IR physics by pure abelian gauge theory breaks down, and has
to be replaced by something more complicated.  Correspondingly, the complex integrable system $\cI'$ generally gets extended by adding
some singular fibers (degenerations of tori) over $u \in \bsing$.  Altogether we get a complete holomorphic symplectic manifold $\cI$ fibered over the whole $\cB$.

\subsection{Compactification of $\N=2$ theories on $S^1$} \label{sec:general-compactification}

\begin{shaded}
Compactifying on $S^1$ turns the integrable system into an honest \hk space.
\end{shaded}

In \S\ref{sec:integrable-system} we have reviewed the
complex integrable system $\cI$ which governs the IR physics of the four-dimensional
field theory $T$.  In that discussion $\cI$ appeared in a somewhat indirect way.
Now we describe a way to see $\cI$ more directly.

Compactify $T$ on $S^1$ of length $2 \pi R$.
At energies $E \ll 1/R$ the resulting physics should be described by a three-dimensional field theory $T[R]$.
To get a first approximation to the physics of $T[R]$, we can consider the dimensional reduction of the local IR Lagrangians
describing $T$ (at least if we stay away from $\bsing$).
Then the fields will be as follows:  $r$ complex scalars, $r$ abelian gauge fields, and $r$ periodic real
scalars (the holonomies of the gauge fields around $S^1$.)  We can moreover dualize the abelian gauge fields to get another
$r$ periodic scalars, so altogether we have $r$ complex scalars and $2r$ periodic real scalars.  The complex scalars
parameterize a sigma model into $\cB$, and we can think of the $2r$ periodic
real scalars as giving a map into a $2r$-torus; so locally we now have
a sigma model into a product of $\cB$ with a real $2r$-torus.

To find the global structure of this sigma model, one has to
keep track of the EM duality transformations needed to glue together the various local IR Lagrangians of $T$.  After so doing,
one finds that $T[R]$ is a sigma model whose target is
the complex integrable system $\cI'$ which we described in \S\ref{sec:integrable-system}.
Thus, after compactification the integrable system ``comes to life.''

We should clarify the meaning of the statement that $T[R]$ is a sigma model into $\cI'$.
In \S\ref{sec:integrable-system} we described
$\cI'$ only as a holomorphic symplectic manifold.  Now we are getting an actual
sigma model into a Riemannian manifold $\cM'[R]$.
Thus we should ask:  how is the Riemannian manifold $\cM'[R]$ related to the holomorphic symplectic manifold $\cI'$?

The answer is as follows.
The constraints of $\N=4$ supersymmetry in 3 dimensions dictate that the metric on $\cM'[R]$ must be \hk \cite{AlvarezGaume:1981hm,Seiberg:1996nz}.
Since $\cM'[R]$ is \hk, it carries a family of complex structures $J_\zeta$, parameterized by
$\zeta \in \C\PP^1$, as well as corresponding holomorphic symplectic forms $\varpi_\zeta$.
One of these complex structures, $J_0$, is distinguished.  When considered as a holomorphic symplectic manifold
in the complex structure $J_0$, $\cM'[R]$ is identical to $\cI'$.

The exact IR physics (as opposed to the physics obtained by naive dimensional reduction) is also given by a sigma model into $\cM'$.
However, the exact \hk metric on $\cM'[R]$ is \ti{not} the same as the one obtained by naive dimensional reduction:
rather they
differ by quantum corrections which can be computed in terms of the spectrum of BPS particles of $T_4$ \cite{Seiberg:1996nz,Hanany:2000fw,Gaiotto:2008cd}.
The corrections due to a BPS particle of mass $M$
go like $e^{-R M}$ in the limit $R \to \infty$, so in this limit the two metrics converge to one another uniformly,
\ti{except} around points where the mass of some BPS particle goes to zero.  The locus where this happens is precisely
$\bsing$, so around $\bsing$ the quantum corrections are not suppressed even in the $R \to \infty$ limit.\footnote{The fact that
quantum correction are not suppressed around $\bsing$
is a good thing: exactly at this locus the naive metric becomes singular, and the quantum corrections smooth out these singularities,
in such a way that the exact corrected metric extends to a complete space $\cM[R]$ which includes fibers over $\bsing$.
This smoothing requires a correction which is of
order 1, not suppressed in $R$.}

Although the quantum corrections change the metric on $\cM[R]$, they do not change what the space looks
like in complex structure $J_0$:  even after the corrections, it is still identical to the complex integrable system $\cI$
from \S\ref{sec:integrable-system} \cite{Seiberg:1996nz}.

\subsection{Holomorphic functions and line defects} \label{sec:holomorphic-functions}

\begin{shaded}
Vevs of line defects are global holomorphic functions on the \hk space.
\end{shaded}

The family of complex structures $J_\zeta$ on $\cM[R]$ (parameterized by $\zeta \in \C\PP^1$)
corresponds to a family of 1/2-BPS subalgebras
$\cA_\zeta$ of the $\N=4$ supersymmetry algebra.
Vevs of 1/2-BPS local operators $\cO$ preserving $\cA_\zeta$ thus give $J_\zeta$-holomorphic
functions on $\cM[R]$.

For example, the complex scalars $\cO$ which descend from the vector multiplets in the original theory $T$
preserve $\cA_0$.  It follows that the vevs of these complex scalars are $J_0$-holomorphic functions on $\cM[R]$.
Said otherwise,
the projection $\cM[R] \to \cB$ is a $J_0$-holomorphic map.  Of course this is just what we expect, since we have already said
that in complex structure $J_0$, $\cM[R]$ is the complex integrable system $\cI$, with base $\cB$.

The four-dimensional origin of operators $\cO$ preserving the other subalgebras $\cA_\zeta$, $\zeta \in \C^\times$, is a bit different:
we consider 1/2-BPS \ti{line defects} in the original theory $T$.  Such line defects can preserve various different subalgebras
of the 4-dimensional supersymmery.  Upon circle compactification, the line defects reduce to point operators,
and their preserved subalgebras reduce to the various 1/2-BPS subalgebras
$\cA_\zeta$.
Thus, the vevs of supersymmetric line defects wrapped on $S^1$ are $J_\zeta$-holomorphic functions on $\cM[R]$.

Among the supersymmetric line defects there is a distinguished subset of ``simple'' defects,
characterized by the property that a simple defect is not expressible (in correlation functions) as a 
nontrivial sum of other defects.  We expect that every defect can be uniquely decomposed as a sum of simple 
defects (though this statement is not entirely trivial --- see \cite{Gaiotto:2011tf, Cordova2013} 
for more discussion.)

The existence of simple line defects implies in particular that there should be a distinguished vector space 
basis of the space of $J_\zeta$-holomorphic functions on $\cM[R]$.
Distinguished bases for coordinate rings of various algebraic spaces 
(and their quantum deformations)
have been studied in Lie theory (following pioneering work of Lusztig, e.g. \cite{MR1035415}) 
and more generally in algebraic geometry (see e.g. \cite{Gross2012}).  Indeed,
the investigation of these ``canonical bases'' was an important motivation
for the theory of cluster algebras \cite{MR1887642}.  On the other hand, it has turned out 
independently that cluster algebras are closely related to the algebras of line defects
(see e.g. \cite{Gaiotto:2010be,Cecotti:2010fi,Xie2012,Xie2013} for more on this.)
Thus it seems natural to suspect that the canonical bases studied in mathematics 
can be identified with the ones coming from simple line defects.
This point remains to be understood more precisely.

\subsection{Hyperholomorphic bundles and surface defects} \label{sec:hyperholomorphic-bundles}

\begin{shaded}
Surface defects give hyperholomorphic bundles on the \hk space.
\end{shaded}

Now, as in \cite{Gukov2007,Alday2010a,Gaiotto:2009fs,Gaiotto:2011tf}, 
let us consider 1/2-BPS \ti{surface defects} in the four-dimensional theory $T$.  We focus
on defects which are massive in the IR, with finitely many vacua.  Let $\bS$ be such a defect.
Upon compactification of both $T$ and $\bS$ on $S^1$,
$\bS$ has a finite-dimensional Hilbert space of ground states, which we denote $V(\bS)$.

To be more precise, the Hilbert space $V(\bS)$ actually depends on which vacuum of the theory $T[R]$ we are in.
Thus we have a family of Hilbert spaces varying over the moduli space $\cM[R]$.  Said otherwise, $V(\bS)$ is
a Hermitian vector bundle over $\cM$.  The supersymmetry in the situation implies that this vector bundle
is \ti{hyperholomorphic}:  in particular, it admits a family of holomorphic structures, one for each $\zeta \in \C\PP^1$,
compatible with the family of underlying complex structures $J_\zeta$ on $\cM[R]$.

Suppose given two such defects, labeled $\bS$ and $\bS'$, and a 1/2-BPS \ti{interface} between them.
As observed in \cite{Gukov:2006jk}, such an interface can be viewed as a kind of line defect which is restricted to live on the surface defect
rather than roaming free in the 4-dimensional bulk.
Upon circle compactification, this picture reduces to a pair of line defects separated by a local operator.
The local operator preserves $\cA_\zeta$ for some $\zeta$ (just as in the case with no surface defects).
The vev of this local operator is then a $J_\zeta$-holomorphic section of $\Hom(V(\bS), V(\bS'))$.

\subsection{Line defects in the IR}

\begin{shaded}
UV line defects can be expanded in terms of IR ones; the coefficients of this expansion are integers which jump as parameters are varied.
\end{shaded}

In \S\ref{sec:holomorphic-functions} we considered line defects from the UV perspective.  On the other hand,
we could also consider the theory in the IR, in the vacuum labeled by some $u \in \cB$.  As we recalled in \S\ref{sec:integrable-system},
the IR physics is governed by \ti{abelian} $\N=2$ gauge theory.
In pure abelian gauge theory, for any $\zeta \in \C^\times$
we can concretely describe the full set of simple $\zeta$-supersymmetric line defects:
for every $\gamma$ in the EM charge lattice $\Gamma_u$, there is a $\zeta$-supersymmetric abelian
Wilson-'t Hooft operator $L(\gamma)$.

Now, given a $\zeta$-supersymmetric line defect $L_{UV}$,
we can ask how the same defect appears in the IR.  It will look like some integer linear combination
of the simple defects of the IR theory:
\begin{equation} \label{eq:uv-ir}
L_{UV} \leadsto \sum_{\gamma \in \Gamma_u} \fro(L_{UV}, \gamma) L_{IR}(\gamma)
\end{equation}
The coefficients $\fro(L_{UV}, \gamma) \in \Z$ of this expansion can be interpreted as
indices counting supersymmetric ground states of the theory with $L_{UV}$ inserted at some fixed
spatial point, extended in the time direction.  These
states were called \ti{framed BPS states} in \cite{Gaiotto:2010be}.

Importantly, the $\fro(L_{UV}, \gamma)$ can jump as we vary the parameters $(u, \zeta)$: this is the phenomenon of (framed)
wall-crossing.  
The jumps occur when a framed BPS bound state decays or forms, by binding or releasing an unframed BPS
state; thus the precise way in which the $\fro(L_{UV}, \gamma)$ jump is determined by the (unframed) BPS degeneracies of the
theory.  Indeed, studying the jumps of the $\fro(L_{UV}, \gamma)$ gives a lot of information
about the unframed BPS degeneracies:  in particular, it is one way
of establishing that these degeneracies obey the 
celebrated Kontsevich-Soibelman wall-crossing formula \cite{ks1}.

Now, let us again consider compactifying on $S^1$ and taking vevs.  Then \eqref{eq:uv-ir} becomes an equation relating
the vev of $L_{UV}$ to a sum of vevs of defects $L_{IR}$:
\begin{equation} \label{eq:vev-expand}
\langle L_{UV} \rangle = \sum_{\gamma \in \Gamma_u} \fro(L_{UV}, \gamma) \langle L_{IR}(\gamma) \rangle
\end{equation}
However, the quantities $\langle L_{IR}(\gamma) \rangle$ are not
as simple as they would be in the pure abelian gauge theory --- they are significantly corrected by contributions from higher-dimension operators.  
Indeed, to get an indication of how subtle these quantities are, note that
$\IP{L_{UV}}$ should be continuous as a function of
the parameters $(u, \zeta)$ (since the UV theory $T$ has no phase transition), 
while we have just said that the coefficients $\fro(L_{UV},\gamma)$
jump at some walls in the $(u, \zeta)$ parameter space.\footnote{These walls are known by various names:  ``BPS walls'' in 
\cite{Gaiotto:2010be}, 
``$\cK$-walls'' in \cite{Gaiotto2012}, ``walls of second kind'' in \cite{ks1}, or parts of the ``scattering diagram'' in \cite{Gross2011a}.}  Thus, for \eqref{eq:vev-expand} to be consistent, the vevs
$\IP{L_{IR}(\gamma)}$ must also jump at these $\cK$-walls.  As with $\fro(L_{UV}, \gamma)$, 
the jumps of $\IP{L_{IR}(\gamma)}$ are completely determined by the unframed BPS degeneracies.

We will return to the meaning of \eqref{eq:vev-expand} when we consider theories of class $S$, below.

\subsection{Asymptotics} \label{sec:asymptotics}

\begin{shaded} Vevs of line defects are \ti{asymptotically} related to functions on the Coulomb branch. \end{shaded}

For each $\zeta \in \C^\times$, the $\zeta$-supersymmetric 
IR line defect vev $\IP{L_{IR}(\gamma)}$ is a $J_\zeta$-holomorphic function on $\cM[R]$.  
These functions have an important asymptotic property:
as $\zeta \to 0$, they behave as \cite{Gaiotto:2008cd,Gaiotto:2010be}
\begin{equation} \label{eq:asymp}
\IP{L_{IR}(\gamma)} \sim c(\gamma) \exp(\zeta^{-1} \pi R Z_\gamma)
\end{equation}
where $c(\gamma)$ is some $\zeta$-independent constant, and $Z_\gamma$
is the central charge function, pulled back from the Coulomb branch $\cB$.

These asymptotics are realized in a rather nontrivial way.  As we have emphasized, the $\IP{L_{IR}(\gamma)}$
are not continuous, but have jumps corresponding to BPS states of the theory $T$.  
If we fix a point of $\cM[R]$ and look only at the $\zeta$ dependence, then the 
loci where the jumps occur are rays in the $\zeta$-plane, all of which run into the origin.  
These jumps however do not destroy the asymptotics --- rather the discontinuity across each ray becomes
trivial in the $\zeta \to 0$ limit.  This is an example of the Stokes phenomenon.

One concrete consequence
is that the expansion of each $\IP{L_{IR}(\gamma)}$ around $\zeta = 0$ will be given only by an \ti{asymptotic}
series, not a convergent one (a convergent series would necessarily converge to a continuous function,
but $\IP{L_{IR}(\gamma)}$ is not continuous in any disc around $\zeta = 0$.)

\section{Theories of class $S$ and Hitchin systems}\label{sec:class-s}

In this section we specialize from general theories $T$ to theories of class $S$, $T = S[\fg, C]$.
These are theories obtained by compactification of the $(2,0)$ theory from 6 to 4 dimensions;
for the definition see \cite{Gaiotto:2009hg,Gaiotto:2009we},
or \cite{internal-ref} in this volume.
In these theories we will see the role of the Hitchin system.

\subsection{Theories of class $S$}

\begin{shaded} For theories of class $S$, the \hk manifold which appears upon 
compactification to three dimensions is a Hitchin system. \end{shaded}

Now suppose that $T$ is a theory of class $S$, $T = S[\fg, C]$.
The general discussion of \S\ref{sec:general-compactification} applies to this particular theory.
Thus compactifying $T$ on $S^1$ gives a sigma model $T[R]$ into an \hk manifold
$\cM$.  In this case, we can understand concretely what $\cM$ is, as follows.

The 3-dimensional theory $T[R]$ has several descriptions summarized in this picture (arrows mean
``compactify and take IR limit''):
\begin{equation}
\begin{tikzcd}
{} & S[\fg] \arrow{ldd}{C} \arrow{rd}{S^1_R} \arrow{ddd}{C \times S^1_R} & \\
{} & & \text{5-d $\fg$-super-Yang-Mills} \arrow{ldd}{C} \\
 T = S[\fg, C] \arrow{rd}{S^1_R} & & \\
{} & T[R] &
\end{tikzcd}
\end{equation}
The left side of the picture is how we have described $T[R]$ up to now:  $T[R]$ is the IR limit of the compactification of $S[\fg,C]$
on $S^1_R$, and $S[\fg, C]$ in turn can be understood as the IR limit of the compactification of the six-dimensional theory
$S[\fg]$ on $C$.  Altogether this means that $T[R]$ is simply the IR limit of the compactification of $S[\fg]$ on $C \times S^1$,
as indicated by the middle arrow of the picture.
Finally we may do this compactification in the opposite order, obtaining the right side of the picture.  We first compactify $S[\fg]$
on $S^1_R$ and take an IR limit to obtain 5-dimensional super Yang-Mills with gauge algebra 
$\fg$.\footnote{Actually, specifying $\fg$ does not quite determine the 5-dimensional theory; for that we should really specify a particular Lie group $G$ with Lie algebra $\fg$.  Which $G$ we get depends on a subtle
discrete choice which appears upon compactification, as described e.g. in \cite{Freed2012},
using some subtleties of the 6-dimensional $S[\fg]$ explained in \cite{Witten:2009at}.}
Then we compactify this 5-dimensional super Yang-Mills theory on $C$ and take an IR limit to get $T[R]$.
This leads to the statement that $T[R]$ is a sigma model into the moduli space of vacuum configurations of 5d super Yang-Mills
on $C \times \R^{2,1}$ which are translation invariant in the $\R^{2,1}$ directions.

The requirement of translation invariance along $\R^{2,1}$ means that the BPS equations on $C \times \R^{2,1}$
reduce to equations for fields on $C$.
These equations turn out to be some celebrated equations in gauge theory:
they are the Hitchin equations \eqref{hitchin-eq}, which we discuss in \S\ref{sec:harmonic-bundles} below.
(More precisely, the equations which appear are the Hitchin equations modified by the rescaling
$\varphi \to R \varphi$.)  This was essentially observed in
\cite{Bershadsky:1995vm,Harvey:1995tg} (in a slightly different context, but the mathematical problem is the same); see also \cite{Cherkis:2000ft,Cherkis:2001gm} where some important 
special cases were rediscovered in a context closer to ours.

Thus the target $\cM[R]$ of the sigma model $T[R]$ is the moduli space of solutions of Hitchin equations.
For the moment we do not need the detailed form of these equations:  we will just need a few basic properties of $\cM[R]$.
In particular,
\begin{itemize}
\item $\cM[R]$ is a \hk space (\S\ref{sec:harmonic-hyperkahler}), as required by $\N=4$ supersymmetry
in three dimensions.
\item In its complex structure $J_0$, $\cM[R]$ can be identified with
a complex integrable system $\cI$ (\S\ref{sec:hitchin-fibration}), as expected
from \S\ref{sec:integrable-system}-\ref{sec:general-compactification}.
\end{itemize}

Let us say a bit more about this integrable system, specializing to the case $\fg = A_{K-1}$ for
concreteness.
\begin{itemize}
 \item The base of the integrable system $\cI$ is the ``Hitchin base'' (\S\ref{sec:hitchin-fibration}).
On the other hand, from \S\ref{sec:integrable-system} we know that the base should be
the Coulomb branch of $T$.  Thus the Coulomb branch $\cB$ of $T$ can be identified with the Hitchin base.
In particular, the points $u \in \cB$ correspond to algebraic curves $\Sigma_u \subset T^* C$ which are $K$-fold covers of $C$, better known as ``Seiberg-Witten curves.''
 \item The torus fibers $\cI_u$ have a concrete algebro-geometric meaning in terms of the Seiberg-Witten curves $\Sigma_u$,
 as follows:  a point of $\cI_u$ corresponds to 
 a holomorphic line bundle $\cL$ over $\Sigma_u$, with the extra property that
 the determinant of the pushforward bundle
$\pi_* \cL$ is trivial, where $\pi: \Sigma_u \to C$ denotes the covering map
(\S\ref{sec:hitchin-fibration}).
\end{itemize}

\subsection{Line defects} \label{sec:s-line-defects}

\begin{shaded} In the theory $S[A_1, C]$, vevs of line defects are holonomies of flat connections along $C$. \end{shaded}

In \S\ref{sec:holomorphic-functions}
we explained that for any $\N=2$ theory the vevs of $\zeta$-supersymmetric
line defects compactified on $S^1$ should give $J_\zeta$-holomorphic
functions on $\cM[R]$.  In theories of class $S$, these functions turns out to be something quite concrete
and understandable in terms of the curve $C$, as follows.

We will need one more fact about $\cM[R]$ (reviewed in \S\ref{sec:moduli-spaces} below):
$\cM[R]$
is diffeomorphic to the moduli space $\cM_{flat}$ of flat $G_\C$-connections,
via a map $f_\zeta$ which is $J_\zeta$-holomorphic,
\begin{align}
 \cM & \overset{f_\zeta}{\longrightarrow} \cM_{flat} \\
  x  & \mapsto \nabla(x, \zeta)
\end{align}
Thus, if we fix a holomorphic function $F$ on $\cM_{flat}$, we can get a $J_\zeta$-holomorphic function
$F_\zeta$ on $\cM$ by pullback:
\begin{equation} \label{pullback}
F_\zeta(x) = F(\nabla(x,\zeta)).
\end{equation}
The vevs of $\zeta$-supersymmetric
holomorphic line defects arise in this way:  each type of simple line defect $L$ corresponds to
some holomorphic function $F = F_L$ on $\cM_{flat}$.

What are the functions $F_L$ concretely?
Let us restrict our attention to the case $\fg = A_1$.
In these theories we have a complete
understanding of the set of supersymmetric line defects following \cite{Drukker:2009tz,Gaiotto:2010be}.  
The story is especially simple if $C$ has only regular punctures.
In that case, for any $\zeta \in \C^\times$, there are simple $\zeta$-supersymmetric line defects corresponding
to pairs $\{ (\wp, a) \}$, where $\wp$ is a non-self-intersecting
closed curve on $C$, and $a$ a nonnegative integer:
\begin{equation} \label{eq:s-line-correspondence}
 L \leftrightarrow (\wp, a).
\end{equation}
The corresponding function $F_L$ on $\cM_{flat}$ is
\begin{equation} \label{eq:s-line-vev}
 F_L (\nabla) = \Tr \left( {\mathrm {P}}_\nabla(\wp, a) \right),
\end{equation}
where ${\mathrm {P}}_\nabla(\wp, a)$ means the parallel transport of the connection $\nabla$ around the path $\wp$,
in the $(a+1)$-dimensional representation of $SL(2,\C)$.\footnote{Slightly more generally, there are also simple line defects
corresponding to mutually non-intersecting  \ti{collections} of closed curves on $C$, with nonnegative integer weights; the vev
of such a defect is simply the product of the traces associated to the individual curves in the collection.}


For general $\fg$ it seems very likely that there are line defects whose vevs 
give holonomies of $SL(K,\C)$ connections along closed paths, as well as defects corresponding to
more general ``spin networks''; however, the story has not yet been completely developed, and in particular 
it is not yet known how to describe a complete set of
simple line defects.  Some examples have very recently been worked out in \cite{Xie2013}; see also
\cite{Le2012,Sikora2011} for related mathematical work.

\subsection{Interfaces between surface defects} \label{sec:s-interfaces}

\begin{shaded} In the theory $S[A_1, C]$, interfaces between surface defects correspond to parallel transport of flat connections along 
open paths on $C$. \end{shaded}

All the discussion of \S\ref{sec:s-line-defects}
has a natural extension where we replace line defects by interfaces between surface defects
\cite{Alday2010a,Gaiotto:2009fs,Gaiotto:2011tf}, as follows.

In the theory $S[A_1, C]$ there is a natural family of surface defects $\bS^a_z$, labeled by an
integer $a > 0$ and a point $z \in C$.
As we have described in \S\ref{sec:hyperholomorphic-bundles}, each such defect corresponds to a hyperholomorphic vector bundle $V(\bS^a_z)$ over $\cM$.
In this case, $V(\bS^a_z)$ is the $a$-th symmetric power of the \ti{universal harmonic bundle}, restricted to $z \in C$ (see \S\ref{sec:universal-bundle}).
In particular, when we view it as a holomorphic bundle in complex structure $J_\zeta$, $V(\bS^a_z)$ is the $a$-th symmetric power of the \ti{universal flat bundle}, restricted to $z \in C$. 

So much for the surface defects by themselves:  how about interfaces between surface defects?
Much like \eqref{eq:s-line-correspondence}, we have a correspondence
\begin{equation}
 L \leftrightarrow (\wp, a)
\end{equation}
where $\wp$ now denotes an \ti{open} path $\wp$ from $z$ to $z'$, and the corresponding $L$ is a
$\zeta$-supersymmetric \ti{interface} between $\bS^a_z$ and $\bS^a_{z'}$. 
The corresponding vev $F_L$ should be a holomorphic section of $\Hom(V(\bS^a_z), V(\bS^a_{z'}))$.
That section is
\begin{equation}
F_L(\nabla) = {\mathrm {P}}_{\nabla}(\wp, a).
\end{equation}
Thus:  in the theory $S[A_1, C]$, vevs of interfaces between surface defects
are parallel transports of $SL(2,\C)$ connections along
open paths on $C$.

Note that giving the operators ${\mathrm {P}}_\nabla(\wp, a)$ for \ti{all} paths $\wp$ is equivalent to
giving the connection $\nabla$ itself.  Thus, for any fixed $\zeta \in \C^\times$,
studying $\zeta$-supersymmetric interfaces between surface defects
in the theory $S[A_1, C]$ is \ti{equivalent} to studying flat $SL(2,\C)$-connections on $C$.
(Indeed, this gives an alternative derivation of the fact that $\cM[R]$ in complex structure
$J_\zeta$ is isomorphic to the moduli space of flat $SL(2,\C)$-connections.)

\subsection{Line defects in the IR} \label{sec:IR-defects}

\begin{shaded} Vevs of IR line defects give local coordinate systems on the Hitchin moduli space;
one can get Fock-Goncharov and Fenchel-Nielsen coordinates in this way. \end{shaded}

We have said in \eqref{pullback}, \eqref{eq:s-line-vev}
that the vevs $\IP{L_{UV}}$ 
are the $f_\zeta$-pullback from $\cM_{flat}$ of some particular holomorphic 
functions, namely the trace 
functions attached to closed paths on $C$.  

Something similar is true for the IR vevs $\IP{L_{IR}(\gamma)}$.  As we have commented,
these functions are not quite globally holomorphic on $\cM[R]$: 
rather they jump at some codimension-1 loci ($\cK$-walls).  However, suppose 
that we initially restrict to a small neighborhood of some initial $u$, 
and then (if we like) extend $\IP{L_{IR}(\gamma)}$ to a larger domain by 
analytic continuation.  In this case we obtain an honest holomorphic 
function $F_\gamma$, defined on some domain in $\cM[R]$.
These holomorphic functions should be regarded as IR analogues of the $\IP{L_{UV}}$
we considered above, and in some respects they are similar:  in particular, they 
are also the $f_\zeta$-pullback of some
holomorphic functions $F_\gamma$ on $\cM_{flat}$.

Precisely what functions $F_\gamma$ we get in this way depends on our choice of an initial $u$, and also on the parameter $\zeta$.
For any fixed choice, considering all $F_\gamma$ at once gives a local
\ti{coordinate system} on $\cM_{flat}$.
In particular, one can obtain in this way both the Fock-Goncharov and complexified 
Fenchel-Nielsen coordinate systems on $\cM_{flat}$ \cite{Gaiotto:2009hg,Hollands2013}.
The Fock-Goncharov coordinates are obtained for generic choices of $(u, \zeta)$
while some special ``real'' $(u, \zeta)$ (related to Strebel differentials on $C$)
give Fenchel-Nielsen.

Incidentally, the coefficients of the expansion \eqref{eq:uv-ir}, i.e. the framed BPS indices,
have a concrete geometric interpretation:  they are counting geometric objects on the
curve $C$, called ``millipedes'' in \cite{Gaiotto:2010be}.

\subsection{(Non)abelianization}

\begin{shaded} Vevs of IR line defects can be viewed as giving flat $\C^\times$-connections over spectral curves. \end{shaded}

There is another way of viewing the vevs of IR line defects.  
The charge lattice $\Gamma_u$ in the theory $S[A_1, C]$ can be described concretely in terms of the Seiberg-Witten curve $\Sigma_u$.  Indeed,
$\Gamma_u$ sits inside $H_1(\Sigma_u, \Z)$.\footnote{To be precise, consider the projection map $\pi_*: H_1(\Sigma_u, \Z) \to H_1(C, \Z)$; the lattice $\Gamma_u$ is the kernel of this projection.}  Thus in the IR we have line defects corresponding to paths on $\Sigma$.
Moreover, there is a $\C^\times$ connection
$\nabla^\ab(\zeta)$ over $\Sigma$, such that the vev of the simple $\zeta$-supersymmetric
line defect corresponding to the homology class $\gamma$ is the holonomy of $\nabla^\ab(\zeta)$
around any path in the homology class $\gamma$.  One can think of $\nabla^\ab(\zeta)$ as an
``abelianization'' of $\nabla(\zeta)$.

This construction can be summarized in a commutative diagram:
\begin{equation}
\begin{tikzcd}
{} & \cM[R] \arrow{ld}[description]{f_\zeta} \arrow{rd}[description]{f^\ab_\zeta} &  \\
\cM_{flat}(SL(2),C) &  & \arrow{ll}[description]{\Psi_\zeta} \cM_{flat}(\C^\times, \Sigma)
\end{tikzcd}
\end{equation}
The left arrow $f_\zeta$ is the ``UV'' map which takes a vacuum
of $T[R]$ to its corresponding $SL(2,\C)$-connection $\nabla(\zeta)$ over $C$.
The right arrow $f^\ab_\zeta$ is the ``IR'' map which takes a vacuum
of $T[R]$ to its corresponding $\C^\times$-connection $\nabla^\ab(\zeta)$ over $\Sigma$.
The two differ by a third map $\Psi_\zeta$, which we call ``nonabelianization'' since
it takes an abelian connection $\nabla^\ab$ over $\Sigma$ to a nonabelian one $\nabla$ over $C$.
From the fact that the framed BPS counts $\fro$ are piecewise constant, it follows that $\Psi_\zeta$ depends in 
a piecewise constant way on $\zeta$, and its jumps are controlled by the BPS spectrum of the 
theory $S[A_1, C]$.

The story is expected to be similar for arbitrary $\fg$;
in particular, the nonabelianization map $\Psi_\zeta$ was described in detail in \cite{Gaiotto2012} for $\fg = gl(K)$.

\subsection{Asymptotics}

\begin{shaded} Fock-Goncharov and Fenchel-Nielsen coordinates have nice asymptotic properties, when 
evaluated along special 1-parameter families of connections coming from points of the Hitchin system. \end{shaded}

Fix a point $x \in \cM[R]$ and some $\zeta_0 \in \C^\times$.  As we have said in \S\ref{sec:s-line-defects},
there is a corresponding real 1-parameter family of connections $\nabla(t) = f_{t\zeta_0}(x)$, 
$t > 0$, i.e. a real path in $\cM_{flat}$.  On the other hand, as we have explained in \S\ref{sec:IR-defects}, 
the choice of $(x,\zeta_0)$
also determines a particular local coordinate system on $\cM_{flat}$ (by taking vevs of $\zeta_0$-supersymmetric IR line
operators, analytically continued from the initial point $x$.)  In particular, these 
may be Fock-Goncharov or Fenchel-Nielsen coordinates on $\cM_{flat}$.

As $t \to 0$, the coordinates $F_\gamma(\nabla(t))$ thus behave according to \eqref{eq:asymp}.
The general statement \eqref{eq:asymp} involves the central charges of the theory, 
but for this particular theory they can be written more concretely:
\begin{equation}
Z_\gamma = \frac{1}{\pi} \int_\gamma \lambda
\end{equation}
for $\lambda$ the Liouville $1$-form on $T^* C$.
Thus \eqref{eq:asymp} becomes
\begin{equation} \label{eq:asymp-2}
F_\gamma(\nabla(t)) \sim c(\gamma) \exp\left(t^{-1} \zeta_0^{-1} R \oint_\gamma \lambda\right).
\end{equation}
Let us make a few remarks about \eqref{eq:asymp-2}:

\begin{itemize}
\item \eqref{eq:asymp-2} is a version of the WKB approximation, applied to the special family of connections
$\nabla(t)$; indeed, $\nabla(t)$ has the form (see \eqref{hitchin-3-term-connection})
\begin{equation}
\nabla(t) = t^{-1} \varphi + \cdots
\end{equation}
where $\zeta_0^{-1} \varphi$ is a $2 \times 2$ 
matrix-valued 1-form on $C$, whose $2$ eigenvalues are the values 
of $R \lambda$ on the $2$ sheets of $\Sigma$.  

\item \eqref{eq:asymp-2} provides a link between
the Fock-Goncharov or Fenchel-Nielsen coordinates and
the periods of the spectral curve.  This link plays some role in the AGT correspondence --- e.g.
for Fenchel-Nielsen coordinates it seems to be used in \cite{Teschner2011}.  
The nature of the link is somewhat nontrivial
(cf. the comments in \S\ref{sec:asymptotics} about Stokes phenomena); this is in some sense to be expected,
since the two objects we are relating are holomorphic in different complex structures 
on the Hitchin space.  I would very much like to know whether these Stokes phenomena have 
some significance in AGT.

\item All of this is expected to generalize to $\fg = gl(K)$, as outlined in \cite{Gaiotto2012}.  The coordinate systems which appear there seem to be more general than Fock-Goncharov or Fenchel-Nielsen.
Presumably it generalizes further to any $\fg$ of ADE type, but this generalization has not yet 
been worked out.
\end{itemize}

\subsection{Operator products and their quantization}

\begin{shaded} Keeping track of spins of framed BPS states leads to a natural quantization of the Hitchin system. \end{shaded}

As we have described above, the vevs of supersymmetric line defects give a natural basis
for the space of $J_\zeta$-holomorphic functions on $\cM[R]$.  The algebra structure on this space
also has a natural meaning in terms of line defects:  it corresponds to the operator product.
Indeed, writing $LL'$ for the operator product between $\zeta$-supersymmetric 
line defects $L$ and $L'$, we have
\begin{equation} \label{eq:ope-vevs}
\IP{LL'} = \IP{L} \IP{L'}.
\end{equation}
In particular, this vev does not depend on the direction in which $L$ approaches $L'$;
this is a consequence of the more general fact that moving $L$ or $L'$ changes it only by a term
which vanishes in $\zeta$-supersymmetric correlators.

There is an interesting deformation of this story, as follows. 
Let $J_3$ be the generator of a spatial $U(1) \subset SO(3)$, and let $I_3$ be a generator
of some $U(1)_R \subset SU(2)_R$.  We are going to make a modification of the quantum field theory $T$, 
which is most convenient to describe in Hamiltonian language:  we 
insert the operator $(-y)^{2(J_3 + I_3)}$ in all correlation functions
(so all correlation functions become functions of the auxiliary parameter $y$, 
and when $y = -1$ we reduce to the original $T$).
The modified theory is still supersymmetric, but now 
line defects can be supersymmetric only if they are inserted along the 
axis $x^1 = x^2 = 0$.  As a result, in computing the operator product of supersymmetric line defects
we are constrained to consider them approaching one another along this axis.  Once again moving the
defects along the line does not affect supersymmetric correlators, but there are now two possible
orderings of the defects along the line, which have no reason to be equivalent.
Thus, at least as far as supersymmetric correlation functions are concerned,
we have a noncommutative (but still associative) deformation of the operator product
of the original theory.  Upon taking vevs, this then induces a corresponding deformation
of the algebra of $J_\zeta$-holomorphic functions on $\cM[R]$.
This deformation has been discussed in various places including
\cite{Gaiotto:2010be,Ito2011,Drukker:2009id} (see also \cite{internal-ref} in this volume),
and essentially also in \cite{MR2233852,MR2567745}.

For IR line defects we can compute directly in the abelian theory to find the simple deformation
\begin{equation} \label{abelian-noncom}
\IP{L_{IR}(\gamma) L_{IR}(\gamma')} = y^{2 \IP{\gamma,\gamma'}} \IP{L_{IR}(\gamma') L_{IR}(\gamma)}
\end{equation}
(this boils down to working out the angular momentum stored in the crossed electromagnetic
fields between two dyons of charges $\gamma$ and $\gamma'$.)
On the other hand, as we have described, the $\IP{L_{IR}(\gamma)}$ are local coordinates on $\cM[R]$;
in fact they are even local \ti{Darboux} coordinates, i.e.
\begin{equation}
\{ L_{IR}(\gamma) , L_{IR}(\gamma') \} = \IP{\gamma, \gamma'} L_{IR}(\gamma) L_{IR}(\gamma').
\end{equation}
Thus \eqref{abelian-noncom} says that the deformation we are considering is a \ti{quantization}
of the Poisson algebra of functions on $\cM[R]$.

The precise deformation \eqref{abelian-noncom} (``quantum torus'') 
had appeared earlier in \cite{ks1,Dimofte:2009bv} in the context of the 
wall-crossing formulas for refined BPS invariants.  Here we are encountering the same 
deformation in our discussion of line defects and their framed BPS states.  
This is not a coincidence; indeed the refined wall-crossing formula can be understood as 
a necessary consistency condition for the wall-crossing of framed BPS states \cite{Gaiotto:2010be}.

In theories of class $S[A_1]$ the operator product and its quantization are 
given by ``skein relations'' like those familiar in Chern-Simons theory (here
for the 3-manifold $C \times \R$).

\section{Basics on the Hitchin system} \label{sec:hitchin-systems}

In this section we present some background on the Hitchin system, without reference to physics.
Fix a compact Riemann surface $C$, and a compact Lie group $G$.

\subsection{Harmonic bundles} \label{sec:harmonic-bundles}

Hitchin's equations \cite{MR89a:32021} are a system of partial differential equations on $C$.
They concern a triple $(E, D, \varphi)$ where
\begin{itemize}
\item $E$ is a $G$-bundle on $C$,
\item $D$ is a $G$-connection in $E$,
\item $\varphi$ is an element of $\Omega^{1}(\End E)$.
\end{itemize}
For example, if $G = SU(K)$, then $E$ can be considered concretely as 
a Hermitian vector bundle of rank $K$, with trivial determinant; 
in a local unitary gauge,
$D$ is of the form $D = \partial + A$; and both $A$ and $\varphi$ are represented by 1-form-valued skew-Hermitian matrices.

The equations are
\begin{subequations} \label{hitchin-eq}
\begin{align}
 F_D - [\varphi, \varphi] &= 0, \label{hitchin-eq-1} \\
 D \varphi &= 0,  \label{hitchin-eq-2} \\
 D \star \varphi &= 0. \label{hitchin-eq-3}
\end{align}
\end{subequations}
Call a triple $(E, D, \varphi)$ obeying these equations a \ti{\hb}.

When $G$ is abelian, these equations are linear and it is relatively easy to describe the harmonic bundles
(it boils down to Hodge theory for 1-forms on the curve $C$).  For $G$ nonabelian, 
harmonic bundles are harder to describe explicitly.
Nevertheless they do exist, as we will discuss below.

\subsection{Higgs bundles and flat bundles}

Given a harmonic bundle,
by ``forgetting'' some of the structure one can obtain either a Higgs bundle or a
flat bundle.  Remarkably, this ``forgetful'' map turns out to be invertible, so that we
can actually reconstruct the harmonic bundle from a Higgs bundle or a flat bundle.
Let us now describe how this works.

Start from a harmonic bundle $(E, D, \varphi)$.
Now suppose we replace $E$ by its complexification, a $G_\C$-bundle $E_\C$.\footnote{The notation $E_\C$ expresses the fact that the gauge group has 
been complexified; to avoid confusion we emphasize that the corresponding associated vector bundles 
do not get complexified.}
For example, when $G = SU(K)$, $E$ is a Hermitian vector bundle of rank $K$,
and passing from $E$ to $E_\C$ corresponds to forgetting the Hermitian metric and remembering only the underlying
complex vector bundle.
Let us also decompose $D$ and $\varphi$ into their $(1,0)$ and $(0,1)$ components:
\begin{equation} \label{four-data}
(D, \varphi) \to (D^{(0,1)}, D^{(1,0)}, \varphi^{(1,0)}, \varphi^{(0,1)}).
\end{equation}

\subsubsection*{Higgs bundles}

Now, suppose that of the four parts \eqref{four-data} we remember only the pair
\begin{equation} \label{higgs-data}
(D^{(0,1)}, \varphi^{(1,0)}).
\end{equation}
Then what do we have?

The operator $D^{(0,1)}$ induces the structure of holomorphic $G_\C$-bundle on $E_\C$ (namely, holomorphic sections
are the ones which are annihilated by $D^{(0,1)}$.)  Let $E_h$ denote $E_\C$ equipped with this holomorphic structure.
The equations \eqref{hitchin-eq-2}-\eqref{hitchin-eq-3} together imply that
$\varphi^{(1,0)}$ is a holomorphic section of $\End(E_h)$.  Let $\phi = \varphi^{(1,0)}$.

Thus, starting from a harmonic bundle,
we have produced a pair $(E_h, \phi)$
where $E_h$ is a holomorphic $G_\C$-bundle and $\phi$ is an $\End(E_h)$-valued holomorphic 1-form.  Such a pair
is called a \ti{Higgs bundle}.

It looks difficult to recover the original harmonic bundle data \eqref{four-data}
just from the Higgs bundle data \eqref{higgs-data}.
If we remembered the underlying $G$-structure,
we could
reconstruct $D^{(1,0)}$ from $D^{(0,1)}$, and $\varphi^{(0,1)}$ from $\varphi^{(1,0)}$, just by taking adjoints.
However, we have forgotten the $G$-structure, so we do not have a notion of adjoint.  Choosing a random
$G$-structure will not do:  this would allow us to construct \ti{some} $(D,\varphi)$,
but there is no reason why Hitchin's equations would be satisfied.

Nevertheless, the remarkable fact \cite{MR89a:32021,MR944577}
is that given a Higgs bundle there is a unique way to find a $G$-structure such that Hitchin's equations are
indeed satisfied!
(Strictly speaking this is not quite true
for every Higgs bundle, but it is almost true:  one only needs to impose an appropriate
condition of ``stability.''  This condition holds for a generic Higgs bundle.)

So altogether we have two inverse constructions:  one trivial forgetful map from harmonic bundles to Higgs bundles, and one
very nontrivial reconstruction map from Higgs bundles to harmonic bundles.

\subsubsection*{Anti-Higgs bundles}

All of what we have just said has a conjugate version, where instead of \eqref{higgs-data}
we remember only the pair
\begin{equation} \label{anti-higgs-data}
(D^{(1,0)}, \varphi^{(0,1)}).
\end{equation}
These give directly the antiholomorphic version of a Higgs bundle,
which we might call an \ti{anti-Higgs bundle}.  Just as above, we have a forgetful map from harmonic bundles
to anti-Higgs bundles, and an inverse reconstruction map from anti-Higgs bundles to harmonic bundles.
Complex conjugation exchanges Higgs and anti-Higgs bundles, in a way that commutes with all the above maps.

\subsubsection*{Flat bundles}

Now suppose instead that we remember a more interesting combination of the data \eqref{four-data}:
fix some $\zeta \in \C^\times$, and remember only the pair
\begin{equation} \label{connection-data}
(D^{(0,1)} + \zeta \varphi^{(0,1)}, D^{(1,0)} + \zeta^{-1} \varphi^{(1,0)}).
\end{equation}
We may regard these two pieces as the two halves of a complex connection $\nabla$ in $E_\C$:
\begin{equation} \label{hitchin-3-term-connection}
\nabla = \zeta^{-1} \varphi^{(1,0)} + D + \zeta \varphi^{(0,1)}.
\end{equation}
From \eqref{hitchin-eq} it follows that $\nabla$ is flat.
Thus, given a harmonic bundle $(E, D, \varphi)$ and a parameter $\zeta \in \C^\times$, we have obtained a flat bundle $(E_\C, \nabla)$.

Given only the pair $(E_\C, \nabla)$ it is not obvious how to recover the full harmonic bundle $(E, D, \varphi)$.
Nevertheless this can indeed be done, in a unique way \cite{MR887285,MR965220} (again under an appropriate
``stability'' condition, which is generically satisfied).
So the story is parallel to what we said above for Higgs bundles:  we have a trivial forgetful map
from harmonic bundles to flat bundles, and a nontrivial reconstruction map from flat bundles to harmonic bundles.
In fact, here we have a \ti{family} of forgetful and reconstruction maps, parameterized by $\zeta \in \C^\times$.

\subsubsection*{Limits of parameters}

There is a relation between these two constructions, as follows.
Evidently, for any $\zeta \in \C^\times$,
remembering \eqref{connection-data} is equivalent to remembering the pair
\begin{equation} \label{connection-data-rescaled}
(D^{(0,1)} + \zeta \varphi^{(0,1)}, \zeta D^{(1,0)} + \varphi^{(1,0)}).
\end{equation}
In the limit $\zeta \to 0$ this becomes \eqref{higgs-data}.  Thus the
map between harmonic bundles and Higgs bundles is the $\zeta \to 0$ limit of our family of maps between
harmonic bundles and flat connections.
Similarly, the map between harmonic bundles and anti-Higgs bundles arises in the $\zeta \to \infty$ limit.

\subsubsection*{Summing up}

Starting from a harmonic bundle, by ``forgetting'' some information --- in a way depeding on a parameter
$\zeta \in \C\PP^1$ --- we can produce one of three objects:
\begin{enumerate}
\item a Higgs bundle (this arises at $\zeta = 0$),
\item a flat bundle (this arises for any $\zeta \in \C^\times$),
\item an anti-Higgs bundle (this arises at $\zeta = \infty$).
\end{enumerate}

\subsection{Moduli spaces} \label{sec:moduli-spaces}

Now we want to discuss the \ti{moduli space} of harmonic bundles.

Up to equivalence, the $G$-bundle $E$ appearing in the definition of ``harmonic bundle'' is determined by
a discrete topological invariant valued in $\pi_1(G)$.
For example, when $G = PSU(K)$, $E$ is fixed up to equivalence by a single ``Stiefel-Whitney class'' valued in $\Z / K \Z$;
when $G = SU(K)$ all $E$ are equivalent.  Thus it is frequently convenient to fix a single $E$ once and for all.
For the rest of this section let us take this point of view.

Having done so, the remaining equivalences are given by the ``gauge group''
\begin{equation}
\cG = \{\text{smooth sections of } \Aut E\}.
\end{equation}
This $\cG$ has an action on $(D, \varphi)$, under which $D$ transforms as usual for a connection
while $\varphi$ transforms in the adjoint representation.
The equations \eqref{hitchin-eq} are invariant under this action.
In particular, $\cG$ acts on the space of harmonic bundles.

Similarly, the $G_\C$-bundle appearing in the definition of ``Higgs bundles,'' ``flat bundle'' or ``anti-Higgs bundle'' is
determined up to equivalence by
the same discrete topological invariant.  Thus it will be convenient to fix this $G_\C$-bundle to be $E_\C$.
Then the remaining equivalences are given by the ``complexified gauge group''
\begin{equation}
\cG_\C = \{\text{smooth sections of } \Aut E_\C \},
\end{equation}
which thus acts on the space of Higgs bundles, flat bundles or anti-Higgs bundles.

Given two $\cG$-equivalent harmonic bundles, the corresponding Higgs bundles are $\cG_\C$-equivalent,
and vice versa; similarly, given two $\cG$-equivalent harmonic bundles, the corresponding flat bundles
are $\cG_\C$-equivalent, and vice versa.

Now we can describe the equivalences discussed above, at the level of \ti{moduli spaces} (and once
again ignoring stability conditions):

\begin{itemize}
\item Let $\cM = \cM(G,C,E)$ be the moduli space of harmonic bundles modulo $\cG$.
\item Let $\cM_{Higgs} = \cM_{Higgs}(G_\C,C,E_\C)$ be the moduli space of
Higgs bundles modulo $\cG_\C$.
\item Let $\cM_{flat} = \cM_{flat}(G_\C,E_\C)$ be the moduli space of $G_\C$-flat connections modulo $\cG_\C$.\footnote{We drop $C$ here to emphasize that $\cM_{flat}$ can be defined without using the complex structure on $C$, e.g. as the space of representations
$\pi_1(M) \to G_\C$ up to equivalence, although of course it does
still depend on the genus of $C$.}
\item Let $\cM_{anti-Higgs} = \cM_{anti-Higgs} (G_\C,C,E_\C)$ be the moduli space of
anti-Higgs bundles modulo $\cG_\C$.
\end{itemize}

If we choose the topology of $E$ appropriately --- for example if we take $G = PSU(K)$ and take the Stiefel-Whitney class of $E$
to be a generator of $\Z / K\Z$ --- then $\cM$, $\cM_{Higgs}$, $\cM_{flat}$
are actually smooth manifolds.  For a more general choice of $E$
there will be some singularities to deal with, but I will mostly ignore this issue in what follows.

What we have said above
implies that there are diffeomorphisms
$f_0: \cM \to \cM_{Higgs}$, $f_\infty: \cM \to \cM_{anti-Higgs}$,
and a \ti{family}
of diffeomorphisms $f_\zeta: \cM \to \cM_{flat}$ parameterized by $\zeta \in \C^\times$:
\begin{equation}
\begin{tikzcd}
{} & \cM  \arrow{ld}[description]{f_0} \arrow{d}[description]{f_\zeta} \arrow[bend left]{d}[description]{f_\zeta} \arrow[bend right]{d}[description]{f_\zeta} \arrow{rd}[description]{f_\infty} &  \\
\cM_{Higgs} & \cM_{flat} & \cM_{anti-Higgs}
\end{tikzcd}
\end{equation}
In particular, this leads to the
very nontrivial statement that $\cM_{Higgs}$ and $\cM_{flat}$ are actually diffeomorphic (via, say, the map $f_1 \circ f_0^{-1}$).

$\cM_{Higgs}$, $\cM_{flat}$ and $\cM_{anti-Higgs}$ all
carry natural complex structures.  It follows that $\cM$ is also complex,
in many different ways:  for any $\zeta \in \C\PP^1$, the diffeomorphism
$f_\zeta$ endows $\cM$ with a complex structure.
We write $J_\zeta$ for this complex structure on $\cM$.

\subsection{Hyperkahler structure} \label{sec:harmonic-hyperkahler}

So far we have explained that the moduli space $\cM$ of harmonic bundles carries a natural family of complex structures $J_\zeta$,
parameterized by $\zeta \in \C\PP^1$.  This might sound exotic at first encounter, but actually there is a natural ``explanation''
for this family of complex structures:  it comes from the fact that $\cM$ carries a natural \hk metric, as we now explain.

Let us fix a $G$-bundle $E$ as we did above.  Then let $\cC$ denote the space consisting of pairs
$(D, \varphi)$ as in \S\ref{sec:harmonic-bundles},
now \ti{without} imposing the Hitchin equations \eqref{hitchin-eq}.
$\cC$ is an infinite-dimensional affine space, with a natural \hk structure.
Moreover $\cC$ is naturally acted on by the gauge group $\cG$.  This action preserves the \hk structure and has a moment map $\vec{\mu}$;
the Hitchin equations say that the three components of $\vec{\mu}$ vanish.

Thus $\cM = \vec{\mu}^{-1}(0) / \cG$.
But this is precisely the \ti{\hk quotient} $\cC /\!\!/\!\!/ \cG$, as defined in
\cite{Hitchin:1986ea}.  In particular, this implies $\cM$ is \hk \cite{MR89a:32021}.
Now, every \hk manifold carries a canonical family of complex structures
parameterized by $\zeta \in \C\PP^1$, and for our $\cM$, this family is precisely the
family $J_\zeta$ we discussed in \S\ref{sec:moduli-spaces}.

\subsection{Universal bundle} \label{sec:universal-bundle}

A point of $\cM$ corresponds to a harmonic bundle on $C$ up to isomorphism.
It is thus natural to ask whether there is a \ti{universal bundle}, i.e. a bundle $V$ over 
$C \times \cM$ equipped with some geometric structure, which when restricted to
a given $x \in \cM$ gives a harmonic bundle over $C$ in the isomorphism class $x$.  Such a bundle need
not quite exist, but at least it exists up to some twisting (so more precisely it exists 
as a section of a certain gerbe over $C \times \cM$).  Locally on $\cM$ 
we may ignore this twisting, and pretend that we have an honest universal bundle.

For our purposes the most 
important fact about this universal bundle is that it is \ti{hyperholomorphic} \cite{Gaiotto:2011tf}:  
it carries a single unitary connection $D$, whose curvature is of type $(1,1)$ relative to \ti{all}
of the complex structures $J_\zeta$ on $\cM$ (see \cite{MR1919716,MR1486984} for some background
on this notion.)

\subsection{Spectral curves and Hitchin fibration} \label{sec:hitchin-fibration}

The different complex structures on $\cM$ expose different features of the space.  Let us focus for a moment on the complex structure $J_0$.  In this complex structure, as we have explained, $\cM$ is identified with $\cM_{Higgs}$.
One of the fundamental facts about this space is that it is a \ti{complex integrable system}.  In particular, it is a fibration
over a complex base space $\cB$, where the generic fiber is a compact complex torus.

Let us describe where this fibration structure comes from.  To be concrete we will focus on the case where
$G = SU(K)$ or $G = PSU(K)$.

Suppose we are given a Higgs bundle $(E_h, \phi)$.
Then the eigenvalues of $\phi$ in the standard representation of $G$ give a $K$-sheeted branched cover of $C$:
\begin{equation}
\Sigma = \{ (z \in C, \lambda \in T^*_z C):  \det (\phi(z) - \lambda) = 0 \} \subset T^* C.
\end{equation}
$\Sigma$ is the \ti{spectral curve} corresponding to the Higgs bundle $(E_h, \phi)$.
The branch points of the covering $\Sigma \to C$ are those points $z \in C$ where $\phi(z)$ has a repeated eigenvalue.

Now, let $\cB$ be the space of \ti{all} $K$-sheeted branched covers $\Sigma \subset T^*C$ of $C$.
Concretely, $\cB$ is a finite-dimensional complex vector space.
Passing from the Higgs bundle $(E_h, \phi)$ to its spectral curve gives a projection known as the
``Hitchin fibration,''
\begin{equation}
\cM_{Higgs} \to \cB.
\end{equation}
$\cB$ is thus called the ``Hitchin base.''

%

We let $\breg \subset \cB$ be the locus of \ti{smooth} spectral curves, and $\bsing = \cB \setminus \breg$.  $\bsing$ is a divisor
in $\cB$ (discriminant locus).

Now, we have claimed that the 
fibers of $\cM_{Higgs}$ over $\breg$ are complex tori:  where does that come from?  To understand it,
note that a smooth spectral curve $\Sigma$ comes with a tautological holomorphic 
line bundle $\cL$, namely the bundle
whose fiber over $(z, \lambda)$ is the $\lambda$-eigenspace of $\phi(z)$.  Moreover, by pushforward
one can recover the original Higgs bundle $(E_h, \phi)$ from $(\Sigma, \cL)$.

Roughly speaking, then, the fiber of $\cM_{Higgs}$ over a given $\Sigma \in \breg$ 
is the set of all holomorphic line bundles over $\Sigma$, with the correct degree (so that their 
pushforward has the same degree as $E$).  This set is well known to be a compact complex torus.\footnote{This is a consequence of Hodge theory for $(0,1)$-forms on $\Sigma$.  One concrete way of thinking about it, in the case where $\cL$ has degree zero, is that every $\cL$ of degree zero admits a metric for which the Chern connection is flat, and this gives an isomorphism between the set of such $\cL$
and the set of unitary flat connections over $\Sigma$, which is evidently a torus.}
More precisely, this torus is not quite the one we want, because
$\cL$ is not an arbitrary bundle --- it constrained by the requirement that the
pushforward of $\cL$ to $C$ should produce a bundle with trivial determinant.  Thus the correct
statement is that the torus fiber
of $\cM_{Higgs}$ is parameterizing those holomorphic line bundles $\cL$ obeying this constraint.
This torus is known as the Prym variety of the covering $\Sigma \to C$.

\subsection{Allowing singularities}

For the application to gauge theory, it is useful to slightly extend our discussion:
instead of taking $(D, \varphi)$ to be regular everywhere,
we may require them to have singularities at some points of $C$, of some constrained sort.\footnote{These singularities correspond to the \ti{punctures} usually included in the definition of the theories of class $S$.}  Broadly speaking there are two classes of singularity which we might consider:  either
\ti{regular} singularities where the eigenvalues of $\varphi$ have 
only a simple pole, or \ti{irregular} ones where the eigenvalues have singularities of higher order.

Essentially all of the mathematical statements we have reviewed in this section 
have direct extensions to the case with singularities; the main references are
\cite{MR1040197} for the regular case, and \cite{wnh} for the irregular case.

One important point to keep in mind is that in the singular case one has some ``local parameters''
keeping track of the behavior near each singularity:  for example, in the case of a regular singularity,
the local parameters are the residues of the eigenvalues of $\varphi$ and the monodromy of $D$.
In defining the moduli space of harmonic bundles one then has to choose whether to hold these parameters
fixed or let them vary.  If one wants the resulting moduli space to carry a natural \hk metric, 
one should hold them fixed
(morally the reason is that the metric is given by the $L^2$ inner product of the fluctuations, and
variations which change the local parameters around a singularity turn out to be non-normalizable.)

\printbibliography

\end{document}